\documentclass[
superscriptaddress,aps,preprintnumbers,amsmath,amssymb,prd,nofootinbib,twocolumn]{revtex4-1}

\usepackage{graphicx}
\usepackage{epstopdf}
\usepackage{dcolumn}
\usepackage{bm}
\usepackage{hyperref}
\usepackage{color}
\usepackage{amsmath}

\begin{document}


\def\a{\alpha}
\def\b{\beta}
\def\c{\varepsilon}
\def\d{\delta}
\def\e{\epsilon}
\def\f{\phi}
\def\g{\gamma}
\def\h{\theta}
\def\k{\kappa}
\def\l{\lambda}
\def\m{\mu}
\def\n{\nu}
\def\p{\psi}
\def\q{\partial}
\def\r{\rho}
\def\s{\sigma}
\def\t{\tau}
\def\u{\upsilon}
\def\v{\varphi}
\def\w{\omega}
\def\x{\xi}
\def\y{\eta}
\def\z{\zeta}
\def\D{\Delta}
\def\G{\Gamma}
\def\H{\Theta}
\def\L{\Lambda}
\def\F{\Phi}
\def\P{\Psi}
\def\S{\Sigma}

\def\o{\over}
\def\beq{\begin{align}}
\def\eeq{\end{align}}
\newcommand{\gsim}{ \mathop{}_{\textstyle \sim}^{\textstyle >} }
\newcommand{\lsim}{ \mathop{}_{\textstyle \sim}^{\textstyle <} }
\newcommand{\vev}[1]{ \left\langle {#1} \right\rangle }
\newcommand{\bra}[1]{ \langle {#1} | }
\newcommand{\ket}[1]{ | {#1} \rangle }
\newcommand{\EV}{ {\rm eV} }
\newcommand{\KEV}{ {\rm keV} }
\newcommand{\MEV}{ {\rm MeV} }
\newcommand{\GEV}{ {\rm GeV} }
\newcommand{\TEV}{ {\rm TeV} }
\newcommand{\1}{\mbox{1}\hspace{-0.25em}\mbox{l}}
\newcommand{\headline}[1]{\noindent{\bf #1}}
\def\diag{\mathop{\rm diag}\nolimits}
\def\Spin{\mathop{\rm Spin}}
\def\SO{\mathop{\rm SO}}
\def\O{\mathop{\rm O}}
\def\SU{\mathop{\rm SU}}
\def\U{\mathop{\rm U}}
\def\Sp{\mathop{\rm Sp}}
\def\SL{\mathop{\rm SL}}
\def\tr{\mathop{\rm tr}}
\def\mpl{M_{\rm Pl}}

\def\IJMP{Int.~J.~Mod.~Phys. }
\def\MPL{Mod.~Phys.~Lett. }
\def\NP{Nucl.~Phys. }
\def\PL{Phys.~Lett. }
\def\PR{Phys.~Rev. }
\def\PRL{Phys.~Rev.~Lett. }
\def\PTP{Prog.~Theor.~Phys. }
\def\ZP{Z.~Phys. }

\def\dd{\mathrm{d}}
\def\ff{\mathrm{f}}
\def\BH{{\rm BH}}
\def\inf{{\rm inf}}
\def\ev{{\rm evap}}
\def\eq{{\rm eq}}
\def\SM{{\rm sm}}
\def\Mpl{M_{\rm Pl}}
\def\GeV{{\rm GeV}}
\newcommand{\Red}[1]{\textcolor{red}{#1}}
\newcommand{\TL}[1]{\textcolor{blue}{\bf TL: #1}}


\title{
Unified Models of the QCD Axion and Supersymmetry Breaking
}

\author{Keisuke Harigaya}
\affiliation{Department of Physics, University of California, Berkeley, California 94720, USA}
\affiliation{Theoretical Physics Group, Lawrence Berkeley National Laboratory, Berkeley, California 94720, USA}
\author{Jacob M.~Leedom}
\affiliation{Department of Physics, University of California, Berkeley, California 94720, USA}
\affiliation{Theoretical Physics Group, Lawrence Berkeley National Laboratory, Berkeley, California 94720, USA}

\begin{abstract}
Similarities between the gauge meditation of supersymmetry breaking and the QCD axion model suggest that they originate from the same dynamics.
We present a class of models where supersymmetry and the Peccei-Quinn symmetry are simultaneously broken. The messengers that mediate the effects of these symmetry breakings to the Standard Model are identical. Since the axion resides in the supersymmetry breaking sector, the saxion and the axino are heavy. We show constraints on the axion decay constant and the gravitino mass.
\end{abstract}

\date{\today}

\maketitle


\section{Introduction}

One of the most serious problems of the standard model, the so-called  strong CP problem~\cite{'tHooft:1976up,Jackiw:1976pf,Callan:1976je}, is elegantly solved by the Peccei-Quinn (PQ) mechanism~\cite{Peccei:1977hh}.
Another problem, the hierarchy problem, is considerably relaxed by low energy supersymmetry (SUSY)~\cite{MaianiLecture,Veltman:1980mj,Witten:1981nf,Kaul:1981wp}.
The precise gauge coupling unification at a high energy scale also motivates low energy SUSY~\cite{Ellis:1990wk,Amaldi:1991cn,Giunti:1991ta}.

There are several hints for a potential connection between these two physical ideas.
First, models of SUSY breaking often involve spontaneous breaking of global symmetry.
In fact, it is one of the sufficient conditions for SUSY breaking~\cite{Affleck:1983vc}.
It would be illuminating to identify this global symmetry with the PQ symmetry.

Second, if the PQ symmetry breaking field resides in the SUSY breaking sector, the super partners of the axion, namely the saxion and the axino, may obtain large masses~\cite{Carpenter:2009zs,Carpenter:2009sw,Higaki:2011bz,Harigaya:2015soa}. Such a model is free from the cosmological problems associated with light saxions and axinos (see \cite{Kawasaki:2007mk} and references therein).

Finally, one realization of the PQ mechanism, the KSVZ model~\cite{Kim:1979if,Shifman:1979if}, has the following superpotential,
\begin{align}
W = Z Q\bar{Q},
\end{align}
where $Z$ is a PQ charged field with a non-zero vacuum expectation value (VEV), and $Q$ and $\bar{Q}$ are PQ and standard model gauge (especially $SU(3)_c$) charged  fields.
If the chiral field $Z$ also obtains a non-zero $F$ term VEV, the SUSY breaking is mediated to super partners of standard model particles via the gauge interaction. This is nothing but the gauge mediation of SUSY breaking~\cite{Dine:1981za,Dimopoulos:1981au,Dine:1981gu,AlvarezGaume:1981wy,Nappi:1982hm} with messenger fields $Q$ and $\bar{Q}$.

Motivated by these hints, we propose a model where SUSY and the PQ symmetry are simultaneously broken,
and the messenger fields that mediate SUSY breaking and the anomaly of the PQ symmetry are in fact the same.
The model provides a unification for the physics of SUSY breaking and the PQ mechanism.

\section{Unification of SUSY and PQ symmetry breaking}
\label{sec:model}

\subsection{Simultaneous SUSY and PQ symmetry breaking in a single sector}

We introduce chiral fields $M_+$ and $M_-$, whose $U(1)_{\rm PQ}$ charges are $+1$ and $-1$, respectively.
The PQ symmetry is broken by introducing a chiral field $X$ and a superpotential coupling,
\begin{align}
W\supset \kappa X (M_+ M_- -v^2),
\end{align}
where $\kappa$ and $v$ are constants.
SUSY is broken by lifting the flat direction $M_+ M_- = v^2$.
To achieve this, we introduce chiral fields $Z_+$ and $Z_-$, and couple them to $M_\pm$ via mass terms.
The superpotential of this minimal model is then given by
\begin{align}
\label{eq:super1}
W = \kappa X (M_+ M_- -v^2) + \lambda' r v Z_+ M_- + \frac{\lambda'}{r} v Z_- M_+,
\end{align}
where $\lambda'$ and $r$ are constants. By phase rotations of chiral fields, we take all constants in Eq.~(\ref{eq:super1}) to be real.

The simultaneous breaking of the PQ symmetry and SUSY via the superpotential in Eq.~(\ref{eq:super1}) is discussed in~\cite{Carpenter:2009zs,Carpenter:2009sw}.
As is shown in section~\ref{sec:IYIT}, this model is the low energy effective theory of a dynamical SUSY breaking model with a deformed moduli constraint (the IYIT model)~\cite{Izawa:1996pk,Intriligator:1996pu}, and is studied by \cite{Harigaya:2015soa} in the context of the heavy scalar scenario~\cite{Giudice:1998xp,Wells:2003tf,Ibe:2006de,Hall:2011jd,Ibe:2011aa}.
A direct coupling between the SUSY and the PQ breaking sector is also analysed in~\cite{Higaki:2011bz} using an effective field theory.

For $\lambda' < \kappa$, the VEVs of the fields are given by
\begin{align}
\vev{M_+} = r v \sqrt{1 - \frac{\lambda^{'2}}{\kappa^2}},~
\vev{M_-} = \frac{v}{r} \sqrt{1 - \frac{\lambda^{'2}}{\kappa^2}},\nonumber\\
\vev{Z_+} = \vev{Z_-}\equiv z,~
\vev{X} = - \frac{\lambda' z}{\kappa \sqrt{1 - \lambda^{'2} / \kappa^2}},
\end{align}
up to a $U(1)_{\rm PQ}$ rotation. The PQ symmetry is broken by the VEVs of the charged fields $M_{\pm}$ and $Z_{\pm}$, where z is undetermined at tree level. If $\lambda' > \kappa$, the VEVs of $M_\pm$ and $Z_\pm$ vanish, and the PQ symmetry is not broken. Thus we will adopt the above hierarchy and also assume that $\lambda' \ll \kappa$ for simplicity.
SUSY is predominantly broken by the $F$ terms of $Z_\pm$,
\begin{align}
F_{Z_\pm} = - \lambda' v^2.
\end{align}

\subsection{Mass spectrum}
The chiral field $X$ and a linear combination of $M_\pm$ obtain a large mass $\kappa v$. We may integrate them out and parametrize $M_\pm$ as
\begin{align}
\label{eq:meson vev}
M_{+} \rightarrow r v \times {\rm exp}(-\frac{A}{v \sqrt{r^2 + 1/r^2}}),\nonumber \\
M_{-} \rightarrow \frac{v}{r} \times {\rm exp}(\frac{A}{v \sqrt{r^2 + 1/r^2}}),
\end{align}
where $A$ is a chiral field.
The effective superpotential of $Z_\pm$ and $A$ is then given by
\begin{align}
W_{\rm eff} = \lambda f^2 Z_+ {\rm exp}(\frac{A}{ \sqrt{2}f}) + \lambda f^2 Z_- {\rm exp}(-\frac{A}{ \sqrt{2}f }),
\end{align}
where $f \equiv v \sqrt{(r^2 + 1/r^2)/2}$ and $\lambda \equiv 2\lambda' / (r^2 + 1/r^2)$.
We note that most of the following discussion relies only on this effective superpotential, and not on the UV completion in Eq.~(\ref{eq:super1}).

Let us first calculate the masses of scalar components of $Z_\pm$ and $A$.
We decompose scalar components as
\begin{align}
Z_\pm \rightarrow &\left( z + \frac{ \pm \rho_H + \rho_L}{2}\right) {\rm exp}\left( i \frac{\pm \theta_H + \theta_L}{2 z} \right),\nonumber \\
A \rightarrow &\frac{s + i \phi}{\sqrt{2}}.
\end{align}
Expanding the scalar potential, we obtain the mass terms,
\begin{align}
V_{\rm mass} =&
\frac{1}{2}\lambda^2 f^2 \left( \theta_H + \frac{z}{f} \phi \right)^2 \nonumber \\
&+\frac{1}{2}\lambda^2 f^2 \left( \rho_H + \frac{ z}{f} s \right)^2
+ \lambda^2 f^2 s^2.
\end{align}
The mass eigenstates and eigenvalues are given by
\begin{align}
a =& \frac{\phi - \epsilon \theta_H}{\sqrt{1 + \epsilon^2 }},~~
b = \frac{\theta_H + \epsilon \phi}{\sqrt{1 + \epsilon^2}},~~
\epsilon \equiv  \frac{z}{f} \nonumber \\
\begin{pmatrix}
\sigma_+ \\ \sigma_-
\end{pmatrix}
=&
\begin{pmatrix}
{\rm cos}\alpha & -{\rm sin}\alpha\\
{\rm sin}\alpha & {\rm cos}\alpha
\end{pmatrix}
\begin{pmatrix}
s \\ \rho_H
\end{pmatrix},\nonumber \\
{\rm tan}\alpha =& \frac{2\epsilon}{1 + \epsilon^2 + \sqrt{1 + 6 \epsilon^2 + \epsilon^4}},
\nonumber \\
m_a= &0,~~m_b = \lambda f \sqrt{1 +\epsilon^2 },\nonumber \\
m_{\sigma_\pm}^2 =& \frac{1}{2}\lambda^2 f^2 \left[ 3 + \epsilon^2 \pm \sqrt{1 + 6 \epsilon^2 + \epsilon^4} \right].
\end{align}
Scalar fields $\rho_L$ and $\theta_L$ are massless at tree level but obtain masses through quantum corrections, as we will see later. The remaining massless field, $a$, is the axion.

Next we consider the masses of the fermionic components of $Z_\pm$ and $A$.
The quadratic terms of $\delta Z_\pm\equiv Z_\pm-z$ and $A$ in the superpotential in Eq.~(\ref{eq:super1}) are
\begin{align}
W_{\rm eff,quad}= \frac{1}{2}\lambda z A^2 + \lambda f A \frac{1}{\sqrt{2}} (\delta Z_+-\delta Z_-). 
\end{align}
The mass eigenstates $\psi_\pm$ and eigenvalues are
\begin{align}
\begin{pmatrix}
\psi_+ \\ \psi_-
\end{pmatrix}
= &
\begin{pmatrix}
{\rm cos}\beta & -{\rm sin}\beta\\
{\rm sin}\beta & {\rm cos}\beta
\end{pmatrix}
\begin{pmatrix}
\psi_A \\ \psi_{Z_H}
\end{pmatrix},~~
{\rm tan}\beta = \frac{\sqrt{\epsilon^2 +4} -\epsilon}{2}, \nonumber \\
m_{\psi_\pm} = &\frac{1}{2}\lambda f  \times \left[ \sqrt{\epsilon^2 +4} \pm \epsilon  \right],
\end{align}
where $\psi_A$ and $\psi_{Z_H}$ are the fermionic components of $A$ and $Z_H \equiv (Z_+ - Z_-)/\sqrt{2}$, respectively.
The fermionic component of $Z_L \equiv ( Z_+ + Z_-)/\sqrt{2}$ is the goldstino and is eaten by the gravitino via the super Higgs mechanism.

The expressions for the mass eigenstates are simplified in the limit $\epsilon \ll1$ or $\epsilon \gg 1$. In the limit $\epsilon \ll 1$, where the PQ symmetry is dominantly broken by the VEVs of $M_\pm$, we have
\begin{align}
a = \phi,~b=\theta_H,~\sigma_+ = s,~\sigma_- = \rho_H,\\
m_b = \lambda f ,~m_{\sigma_+} = \sqrt{2}\lambda f,~m_{\sigma_-} = \lambda f,\\
\psi_{\pm} = \frac{1}{\sqrt{2}}\left( \psi_A \mp \psi_{Z_H} \right),\\
m_{\psi_\pm} = \lambda f.
\end{align}
In the limit $\epsilon \gg 1$, where the PQ symmetry is dominantly broken by the VEVs $\langle Z_\pm\rangle$, we obtain
\begin{align}
a = -\theta_H,~b=\phi,~\sigma_+ = s,~\sigma_- = \rho_H,\\
m_b = \lambda z,~m_{\sigma_+} = \lambda z ,~m_{\sigma_-} = \frac{\sqrt{2}\lambda f^2}{z},\\
\psi_+ = \psi_A,~\psi_- = \psi_{Z_H}, \\
m_{\psi_+} = \lambda  z,~m_{\psi_-} =\frac{ \lambda f^2}{z},
\end{align}
where the masses of $\sigma_- =\rho_H$ and $\psi_- = \psi_{Z_H}$ are suppressed by the large Majorana masses $\lambda z $ of $\sigma_+ = s$ and $\psi_+= \psi_A$.

\subsection{Sgoldstino potential in the minimal model}
As we have seen, the directions $\rho_L$ and $\theta_L$, which correspond to the sgoldstino components, are massless at tree level.
Accordingly, $z$ is undetermined at tree level.
Here we discuss the stabilization of the sgoldstino in the mimimal model given by Eq.~(\ref{eq:super1}).

Quantum corrections generate a potential for the scalar component of $Z_L\equiv \left( Z_+ + Z_- \right)/\sqrt{2}$,
\begin{align}
\label{eq:VCW_minimal}
\Delta V_{\pm}(Z_L) =
\frac{\lambda^4 f^4}{512\pi^2}\Bigl[
8(1 + \epsilon^2)^2 {\rm ln}(1 +\epsilon^2)
\nonumber \\
+2 (3 + \epsilon^2 + \sqrt{1 + 6 \epsilon^2 + \epsilon^4})^2 {\rm ln} \frac{3 + \epsilon^2 + \sqrt{1 + 6 \epsilon^2 + \epsilon^4}}{2} \Bigr. \nonumber \\
+
2 (3 + \epsilon^2 - \sqrt{1 + 6 \epsilon^2 + \epsilon^4})^2 {\rm ln} \frac{3 + \epsilon^2 - \sqrt{1 + 6 \epsilon^2 + \epsilon^4}}{2}
\nonumber \\ 
-(\epsilon- \sqrt{4 +\epsilon^2})^4 {\rm ln} \frac{(\epsilon - \sqrt{4 + \epsilon^2})^2}{4} \nonumber \\
-(\epsilon + \sqrt{4 +\epsilon^2})^4 {\rm ln} \frac{(\epsilon + \sqrt{4 + \epsilon^2})^2}{4}  \Bigl.\Bigr] \nonumber \\
\simeq 
\left\{
\begin{array}{ll}
\frac{\lambda^4 f^2}{32\pi^2}(2{\rm ln}2-1)|Z_L|^2 &: |Z_L| \lesssim   f \\
\frac{\lambda^4 f^4}{16\pi^2} {\rm ln} \frac{|Z_L|}{f } & : |Z_L| \gtrsim   f,
\end{array}
\right.
\end{align}
where $\epsilon = |Z_+| / f$.

The supergravity effect induces a tadpole term for $Z_L$,~~
\begin{align}
V(Z_L) = \Delta V_{\pm}(Z_L) + (-2\sqrt{2}\lambda f^2 m_{3/2} Z_L + {\rm h.c.}),
\end{align}
where $m_{3/2}$ is the gravitino mass.
We take $m_{3/2}$ to be real by a $U(1)_R$ rotation.
The gravitino mass is related with the magnitude of the SUSY breaking by the (almost) vanishing cosmological constant condition 
\begin{align}
\label{eq:m32}
\sqrt{3}m_{3/2} =|F_{Z_L}|/\mpl = \sqrt{2}\lambda \frac{f^2}{\mpl}.
\end{align}

The tadpole term induces the VEV of $Z_L$ and the messenger scale~\cite{Kitano:2006wz}.
Assuming $|\vev{Z_L}| \lesssim f$, we obtain
\begin{align}
\label{eq:VEV_Z}
\vev{Z_L} =  \frac{64\sqrt{2}\pi^2}{2{\rm ln}2-1}  \frac{m_{3/2}}{\lambda^3} = \frac{128\pi^2}{\sqrt{3}(2{\rm ln}2-1)\lambda^2} \frac{f^2}{\mpl}.
\end{align}
For small $\lambda$, the formula (\ref{eq:VEV_Z}) yields $|\vev{Z_L}|> f$.
In such a parameter region, the potential of $Z_L$ given by the quantum correction is logarithmic, and cannot stabilize $Z_L$ against the tadpole term.
Instead $Z_L$ is stabilized around $\vev{Z_L}\sim \mpl$ by the supergravity effect.
Later, we couple $Z_\pm$ to the messenger field. If $\vev{Z_L}$ is as large as $\mpl$, the gauge mediated soft masses of supersymmetric standard model (SSM) particles are smaller than the gravitino mass. Thus, in the following, we concentrate on the parameter region where $\vev{Z_L} \ll \mpl$.
Then in the minimal model, $\vev{Z_L}$ is at the most $\mathcal{O}(f)$.

\subsection{Simultaneous mediation of SUSY breaking and the anomaly of  the PQ symmetry }

\begin{figure}[tb]
 \begin{center}
  \includegraphics[width=0.8\linewidth]{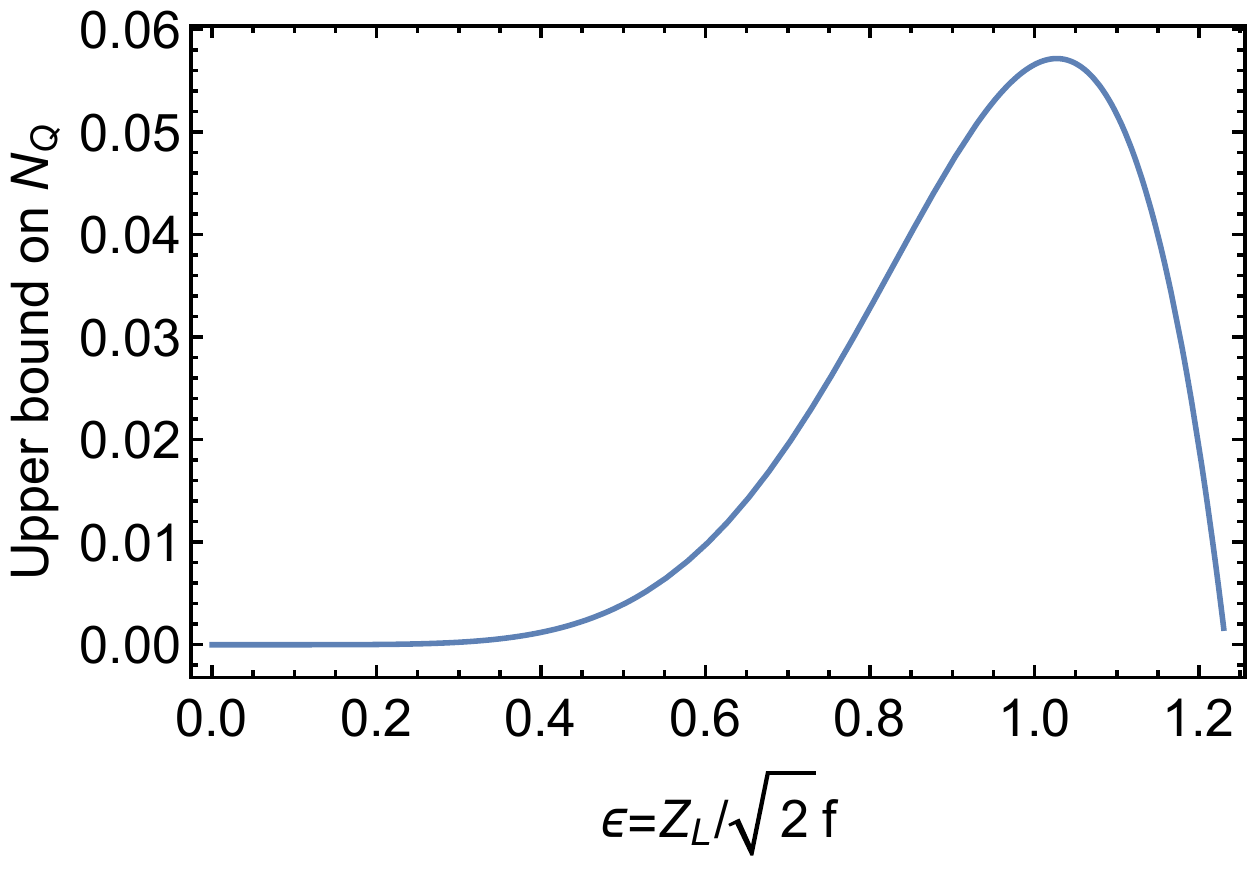}
 \end{center}
\caption{\small\sl 
Upper bound on the multiplicity of the messenger $N_Q$ for the minimal model.
}
\label{fig:NQ_minimal}
\end{figure}

The simplest possibility of the mediation is to introduce a pair of standard $SU(3)_c$ charged chiral fields $Q$ and $\bar{Q}$ with the coupling,
\begin{align}
\label{eq:model1}
W = y Z_+ Q \bar{Q}.
\end{align}
The precise gauge coupling unification is maintained if $Q$ and $\bar{Q}$ are complete multiplets of the $SU(5)$ GUT gauge group.
The mass terms of the scalar component of the messenger field are given by
\begin{align}
V_{\rm mass}=
\begin{pmatrix}
Q^* &  \bar{Q} 
\end{pmatrix}
\begin{pmatrix}
y^2 \vev{Z_+}^2 & yF_{Z_+}^* \\
y F_{Z_+} & y^2 \vev{Z_+}^2
\end{pmatrix}
\begin{pmatrix}
Q \\ \bar{Q}^*
\end{pmatrix}.
\end{align}
To avoid tachyonic masses for the messenger fields, we require that
\begin{align}
\label{eq:y_low}
y >  \frac{|F_{Z_+}|}{ \vev{Z_+}^2} = \frac{2 \lambda f^2 }{\vev{Z_L}^2}.
\end{align}
On the other hand, the quantum correction from the messenger loop generates a potential term for the SUSY breaking field,
\begin{align}
\Delta V_{\rm mes} \simeq  \frac{N_Qy^2}{32\pi^2} F_{Z_L}^2{\rm ln}\frac{|Z_L|^2}{\mu^2},
\end{align}
where $N_Q$ is the multiplicity of the messenger field.
By requiring that this potential does not destabilize the SUSY breaking vacuum, we obtain
\begin{align}
\label{eq:y_high}
\frac{N_Qy^2 \lambda^2 f^4}{8\pi^2 |\vev{Z_L}|^2} < \frac{\partial^2 \Delta V}{\partial |\vev{Z_L}|^2} \equiv 2 m_{Z}^2.
\end{align}
The bounds  on $y$ in Eqs.~(\ref{eq:y_low}) and (\ref{eq:y_high}) are compatible if 
\begin{align}
\label{eq:NQ boound}
N_Q < \frac{4\pi^2  \vev{Z_L}^6}{\lambda^4 f^8} m_{Z}^2 .
\end{align}
In Fig.~\ref{fig:NQ_minimal}, we show the upper bound on $N_Q$ as a function of $Z_L$.
Here we have evaluated $m_{Z}$ using $\Delta V_{\pm}$ in Eq.~(\ref{eq:VCW_minimal}).
It is evident that the upper bound is too severe and is inconsistent with $N_Q \gsim 3$,
which leads us to extend the model to stabilize the sgoldstino.

\subsection{Stabilization of the sgoldstino in extended models: model-independent analysis}

By coupling the sgoldstino to other chiral multiplets, quantum corrections from these multiplets give additional contributions to the mass of the sgoldstino.
Here we assume that a positive squared mass $m_{Z}^2$ is generated from a quantum correction.
(For setups which generate a negative squared mass, see~\cite{Ibe:2010ym,Shih:2007av,Intriligator:2007py,Giveon:2008ne,Evans:2011pz,Curtin:2012yu}.)
Even in this generic situation, we show that there is a lower bound on the axion decay constant and the gravitino mass.

The VEV of $Z_L$ is given by
\begin{align}
\vev{Z_L} =  \frac{4}{\sqrt{3}} \frac{\lambda^2 f^4}{\mpl m_Z^2},
\end{align}
and the gauge mediated gluino mass is given by
\begin{align}
m_{\tilde{g}} = \frac{\alpha_3}{4\pi} \frac{F_{Z_L}}{Z_L} = \frac{\alpha_3}{4\pi} \frac{\sqrt{6}}{4} \frac{ m_Z^2\mpl}{\lambda f^2}.
\end{align}
For given $\lambda$, $f$, and $m_{\tilde{g}}$, $m_Z^2$ is fixed,
\begin{align}
\label{eq:mZ_required}
m_Z^2 = \frac{8\pi}{\alpha_3} \sqrt{\frac{2}{3}} \frac{m_{\tilde{g}}\lambda f^2}{\mpl}.
\end{align}

\begin{figure}[tb]
 \begin{center}
  \includegraphics[width=0.8\linewidth]{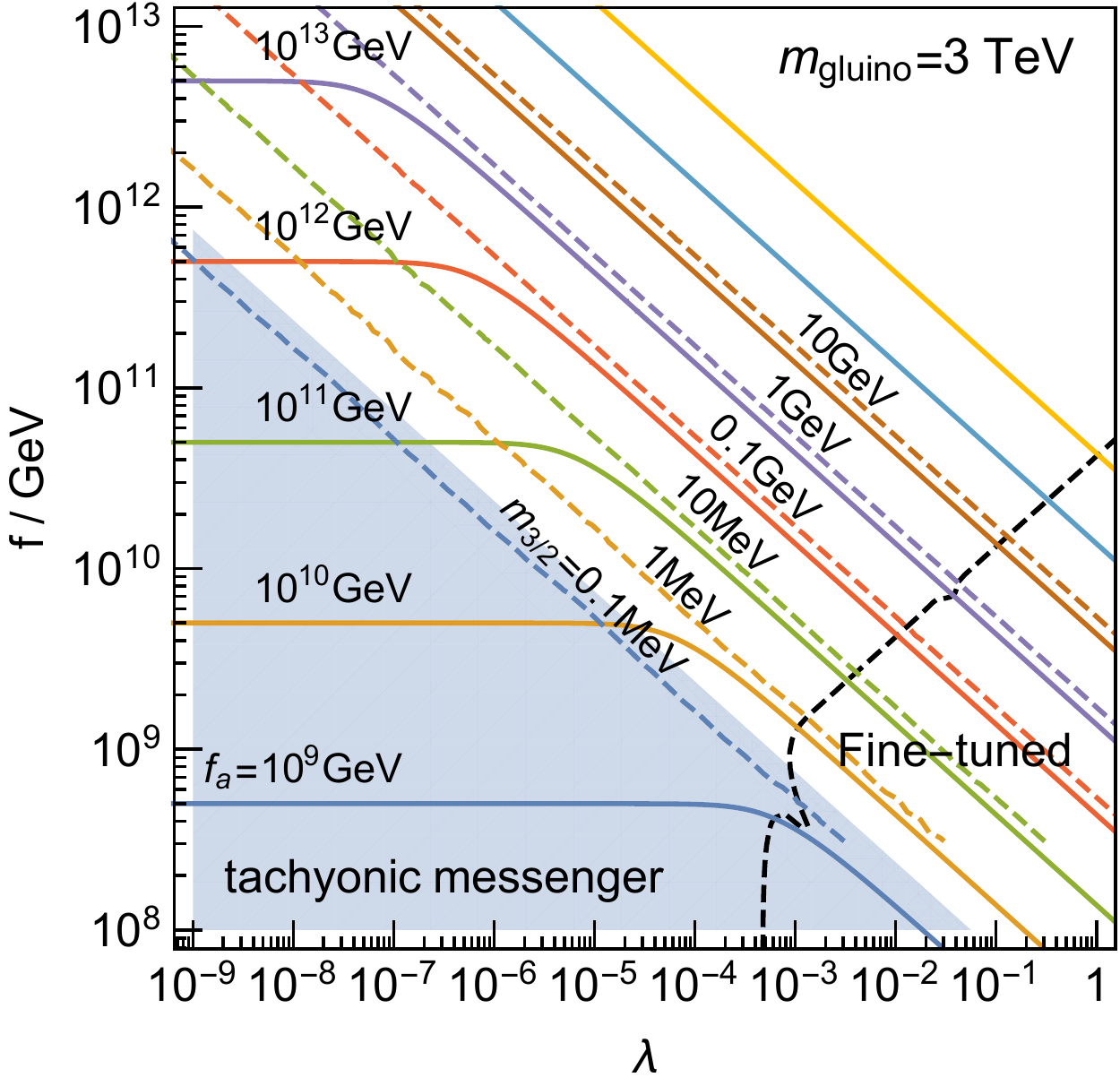}
 \end{center}
\caption{\small\sl 
The model-independent bounds on $(\lambda,f)$ and the contours of the axion decay constant $f_a$ and the gravitino mass $m_{3/2}$.
The blue shaded region is excluded as the messenger field is tachyonic. The region below the black dashed line requires fine-tuning.
}
\label{fig:constraint_ind}
\end{figure}

There are two bounds that must be considered.
One is Eq.~(\ref{eq:NQ boound}),
\begin{align}
N_Q < \frac{\alpha_3^5}{8\sqrt{6} \pi^3} \frac{\lambda^3 f^6}{ m_{\tilde{g}}^5\mpl}.
\end{align}
Another is
\begin{align}
\label{eq:mass bounder}
m_{Z}^2 >  \frac{1}{2}|\frac{\partial^2 \Delta V_\pm}{\partial |\vev{Z_L}|^2}|.
\end{align}
Otherwise we need fine-tuning between $\Delta V_\pm$ and additional contributions to obtain a required value of $m_Z^2$.
In Fig.~\ref{fig:constraint_ind}, we show the constraints on $(\lambda,f)$ as well as the contours of the axion decay constant $f_a$,
\begin{align}
f_a = \sqrt{2 \left(M_+^2 + M_-^2 + Z_+^2 + Z_- ^2 \right)},
\end{align}
and the gravitino mass $m_{3/2}$.
Here we assume that the messenger is in the ${\bf 5}$ representation of the $SU(5)$ GUT group, so $N_Q=5$. For the most part, the axion decay constant is dominated by the VEVs of $M_{\pm}$ in the left half of the parameter space and the VEVs of $Z_{\pm}$ in the right half. The blue shaded region is excluded as the messenger field becomes tachyonic.
The region below a black dashed line calls for fine-tuning.
We obtain lower bounds from them,
\begin{align}
f_a &\gtrsim 1.7\times 10^9 \left(\frac{m_{\tilde{g}}}{\rm 3 TeV}\right)^{2/3}~{\rm GeV},\\
m_{3/2} &\gtrsim 0.2\times\left(\frac{m_{\tilde{g}}}{\rm 3 TeV}\right)^{5/3}~{\rm MeV}.
\end{align}

\subsection{Cosmology}
We now address several cosmological topics that may affect the parameter space of our model.

 Our model contains a SUSY preserving vacuum where the messengers obtain  nonzero VEVs, so we must ensure that the SUSY breaking vacuum is selected during cosmological evolution. Following the discussion in \cite{Fukushima:2012ra}, in the early Universe we assume that the SSM particles are in thermal equilibrium and therefore the sgoldstino field potential obtains finite temperature corrections from the messenger fields. We also take the sgoldstino field to be stabilized at the origin initially due to a positive Hubble-induced mass. The messenger potential becomes unstable about $\langle Z_L\rangle$ = 0 as the universe cools, which causes the  messengers become tachyonic and develop VEVs. To reach the SUSY-breaking vacuum, the sgoldstino field must leave the origin before this occurs, which requires that~\cite{Fukushima:2012ra}
 
\begin{align}
	\frac{y}{\sqrt{2}} < \left(\frac{3^{3/4}}{15}\frac{2g^2+g'^2}{2}\right)^{2/5}\left(\frac{m_{3/2}}{M_{PL}}\right)^{1/5}.
\end{align}
Combining this with Eq.~(\ref{eq:mass bounder}) and Eq.~(\ref{eq:y_low}), we obtain
\begin{align}
\label{eq:cooler}
f_a &\gtrsim 2.6\times 10^{10} \left(\frac{m_{\tilde{g}}}{3 \rm TeV}\right)^{2/3} ~{\rm GeV},\\
 m_{3/2} &\gtrsim 1.6\times \left( \frac{m_{\tilde{g}}}{3 \rm TeV}\right)^{5/3} ~{\rm MeV}.
\end{align}
 Hence vacuum selection raises the lower bounds by a factor $\mathcal{O}$(10).
 
 Another potential concern is that the sgoldstino, which may be produced in the early universe by thermal or nonthermal processes, might affect Big Bang Nucleosynthesis (BBN). The relevant decay modes of the sgoldstino are $Z_L\rightarrow aa$  and $Z_L\rightarrow gg$ with decay rates
 \begin{align}
	\Gamma_{Z_L\rightarrow aa} =& \frac{m_Z^3}{128\pi}\left(\frac{\langle Z_L\rangle}{2f^2+\langle Z_L\rangle^2}\right)^2, \\
	\Gamma_{Z_L\rightarrow gg} =  &\frac{\alpha_3^2}{128\pi^3}\frac{m_z^3}{z^2},
\end{align} 
 respectively.  Looking to the  parameter space in Fig.~\ref{fig:constraint_ind}, the former decay dominates in most of the area where $\langle Z_{\pm}\rangle$ controls the axion decay constant, while the latter decay dominates for a majority of the remaining allowed parameter space.  Sgoldstino decay into gravitinos dominates in the upper right portion of the parameter space but the gravitino is heavy in this region and so it is not favored. In both of the relevant regions, the decay time is short enough that the sgoldstino does not affect BBN.

It should also be noted that the super partners of the axion obtain large masses. This is a merit of the setup described above~\cite{Carpenter:2009zs,Carpenter:2009sw,Higaki:2011bz,Harigaya:2015soa}.
In general, the super partners of the axion obtain only small masses, typically smaller than the masses of SSM particles. Since they couple to SSM particles very weakly while being light, they cause various cosmological problems (see~\cite{Kawasaki:2007mk} and references therein).
These problems are particularly serious in gauge mediation, where the SUSY breaking scale is small and the super partners of the axion are light.
In our setup, since the axion multiplet resides in the SUSY breaking sector, the super partners of the axion can be much heavier than SSM particles and do not cause cosmological problems.

The only light particle that could affect cosmology is the gravitino due to either demanding a low reheating temperature ~\cite{Ellis:1984er,Moroi:1993mb} or overclosing the Universe ~\cite{Pagels:1981ke,Weinberg:1982zq}. The latter issue could potentially be resolved by having the sgoldstino~\cite{Hamaguchi:2009hy,Fukushima:2012ra,Ibe:2006rc}, saxion~\cite{Kim:1992eu,Lyth:1993zw,Kawasaki:2008jc,Co:2016fln}, messenger fields~\cite{Fujii:2002fv} or hidden sector fields~\cite{Ibe:2010ym} dilute the gravitino abundance through large entropy production.

\subsection{Alignment of CP phases}
An interesting feature of our model is that the phases of the gravitino mass and the gaugino masses are aligned with each other.
This is because the phase of the VEV of $\vev{Z_\pm}$, which generates the messenger scale, is aligned with the gravitino mass in a phase convention where the $F$ term of the SUSY breaking field $Z_L$ is real.
Thus, the CP phase of the $B\mu$ term (in a convention where the $\mu$ term is real) due to the supergravity effect~\cite{Moroi:2011fi} is absent in our model.
This feature would be advantageous if one requires that SUSY particles are light (e.g.~to explain the experimental anomaly of the muon anomalous magnetic moment~\cite{Bennett:2006fi,Hagiwara:2011af,Davier:2010nc} by SUSY particles~\cite{Lopez:1993vi,Chattopadhyay:1995ae,Moroi:1995yh}) while the gravitino mass is large (e.g.~to be consistent with a large reheating temperature.)

\section{Example of an extended model: low energy theory of the IYIT model}
\label{sec:IYIT}

\subsection{Effective theory of the IYIT model}
Let us consider a vector-like SUSY breaking sector based on $SU(2)$ hidden strong gauge dynamics~\cite{Izawa:1996pk,Intriligator:1996pu}.
We introduce four chiral fields which are in the fundamental representation of $SU(2)$, $q_i~(i=1\mathchar`-4)$,
and six singlet chiral fields, $Z_+$,~$Z_-$, $Z_{0,a}~(a=1\mathchar`-4)$.
We assume $U(1)_{\rm PQ}$ charges shown in Table~\ref{tab:charge},
and consider the following superpotential,
\begin{align}
\label{eq:super}
W = & \lambda_+ Z_+ q_1 q_2 + \lambda_- Z_- q_3 q_4  \\
&+ Z_{0,a} \left( \lambda_a^{13} q_1 q_3 + \lambda_a^{14} q_1 q_4 + \lambda_a^{23} q_2 q_3 +\lambda_a^{24} q_2 q_4 \right) \nonumber,
\end{align}
where the $\lambda$'s are constants, and summation over $a$ is assumed.
The genericity of the superpotential can be guaranteed by symmetries.
One concrete example of $U(1)_R$ and $Z_4$ charges is shown in Table~\ref{tab:charge}.

\begin{table}[htp]
\caption{Charge assignment of chiral fields}
\begin{center}
\begin{tabular}{|c|c|c|c|c|c|c|c|c|}
& $q_1$ & $q_2$ & $q_3$ & $q_4$ & $Z_+$ & $Z_-$ & $Z_{0,a}$ & $Q\bar{Q}$ \\ \hline
$U(1)_R$ & $0$ & $0$ & $0$ & $0$ & $2$ & $2$ & $2$ & $0$  \\ 
$U(1)_{\rm PQ}$ & $-1/2$ & $-1/2$ & $+1/2$ & $+1/2$ & $1$ & $-1$ & $0$ & $-1$ \\
$Z_4$ & $1$ & $1$ & $1$ & $1$ & $2$ & $2$ & $2$ & $2$
\end{tabular}
\end{center}
\label{tab:charge}
\end{table}

Below the dynamical scale of the hidden $SU(2)$, $\Lambda$,  the theory is described by meson fields $M_{ij} \simeq q_i q_j / \eta \Lambda$ with the deformed moduli constraints, ${\rm Pf} M_{ij} = \Lambda^2 / \eta^2$~\cite{Seiberg:1994bz}.
Here and hereafter, we assume the naive dimensional analysis to count factors of $\eta \sim 4\pi$~\cite{Luty:1997fk,Cohen:1997rt}.
The deformed moduli constraint may be expressed by introducing a Lagrange multiplier field $X$,
\begin{align}
\label{eq:constraint}
W_{\rm eff} = \kappa X \left(M_{12}M_{34} + M_{13}M_{24} + M_{14}M_{23} - \frac{\Lambda^2}{\eta^2}\right).
\end{align}
The tree-level superpotential in Eq.~(\ref{eq:super}) becomes
\begin{align}
W_{\rm tree} =& \lambda_+ \frac{\Lambda}{\eta} Z_+ M_{12} + \lambda_- \frac{\Lambda}{\eta} Z_- M_{34} \\
&  +
\frac{\Lambda}{\eta}  Z_{0,a} \left( \lambda_a^{13} M_{13} + \lambda_a^{14} M_{14} + \lambda_a^{23} M_{23} +\lambda_a^{24} M_{24} \right).\nonumber 
\end{align}
We define
\begin{align}
M_{-} &\equiv M_{12},~M_+\equiv M_{34},\\
M_{0,1}&\equiv \frac{1}{\sqrt{2}} \left(M_{13}+ i M_{24} \right),
M_{0,2}\equiv \frac{1}{\sqrt{2}} \left(M_{13}- i M_{24} \right),\nonumber \\
M_{0,3}&\equiv \frac{1}{\sqrt{2}} \left(M_{14}+ i M_{23} \right),
M_{0,4}\equiv \frac{1}{\sqrt{2}} \left(M_{14}- i M_{23} \right).\nonumber 
\end{align}
Then the effective superpotential in Eq.~(\ref{eq:constraint}) is given by
\begin{align}
W_{\rm eff} = \kappa X \left( M_+ M_- + \frac{1}{2} M_{0,a}^2 - \frac{\Lambda^2}{\eta^2} \right).
\end{align}
By $SU(4)$ rotations of $M_{0,a}$ and $Z_{0,a}$, the total superpotential can be simplified as
\begin{align}
W = &
\kappa X \left( M_+ M_- + \frac{1}{2} c_{ab}M_{0,a}M_{0,b} - \frac{\Lambda^2}{\eta^2} \right)  \\
 & + 
\lambda_+ \frac{\Lambda}{\eta} Z_+ M_{-} + \lambda_- \frac{\Lambda}{\eta} Z_- M_{+} + \lambda_{0,a} \frac{\Lambda}{\eta}  Z_{0,a} M_{0,a}\nonumber
\end{align}
with $c_{ab}$ as a unitary matrix. We will work with only one pair of neutral fields $(Z_0,M_0)$, which corresponds to the generic case that there exists a mild hierarchy in the neutral coupling constants so that the effect of only one neutral field dominates. Therefore, after a redefinition of constants, we have the effective superpotential
\begin{align}
\label{eq:super_extention}
W =& \kappa X (M_+ M_- + \frac{c}{2}M_0^2 -v^2) \nonumber \\
&+ \lambda' r v Z_+ M_- + \lambda' \frac{1}{r} Z_- M_+ + \lambda_0' v Z_0M_0.
\end{align}
The coupling constant $\kappa$ originates from strong dynamics and is expected to be large.
The absolute value of the coupling constant $c$ is at maximum unity.
To maximize the quantum correction, we assume $|c|=1$ in the following.
We also assume that $\lambda_0' v^2 > \lambda f^2$, since otherwise $M_0$ obtains a VEV instead of $M_\pm$. 
The vacuum is  then given by
\begin{align}
\vev{M_+} &=  rv,~\vev{M_-}=\frac{1}{r} v,~\vev{Z_+} = \vev{Z_-} =z,\nonumber \\
 \vev{M_0} &= \vev{Z_0}=0.
\end{align}

\subsection{Stabilization of the sgoldstino by neutral fields in the IYIT model}

To estimate the quantum correction from $Z_0$ and $M_0$, we use the parametrization \cite{Chacko:1998si}
\begin{align}
M_+ \rightarrow r \sqrt{v^2 -  M_0^2/2}, M_- \rightarrow \frac{1}{r} \sqrt{v^2 -  M_0^2/2}.
\end{align}
Here we have neglected the dependence on $A$, which is irrelevant for the quantum correction from $Z_0$ and $M_0$ to $Z_L$.
The effective superpotential of $Z_L$ and $Z_0$, $M_0$ is given by
\begin{align}
\label{eq:Weff2}
W_{\rm eff} \simeq & \lambda f^2 (Z_+ + Z_-) \sqrt{1  - \frac{M_0^2}{2v^2}} + \lambda_0' v Z_0M_0 \nonumber \\
 \simeq & \sqrt{2} \lambda f^2 Z_L - \frac{\sqrt{2}}{4} R^2 \lambda Z_L M_0^2 + \lambda_0 f Z_0 M_0,\nonumber \\
R \equiv & \frac{f}{v} > 1, \lambda_0 \equiv \frac{1}{R} \lambda_0'.
\end{align}

The quantum correction to the potential of $Z_L$ from $Z_0$ and $M_0$ is given by
\begin{align}
\Delta V_{0}
=& \frac{\lambda^4 R^4  f^4 }{64\pi^2} f(\frac{\lambda R^2  z}{\lambda_0 f}) \left( 1 + O\left(\left(\lambda R/\lambda_0\right)^4\right )\right)  \\
\simeq  &
\left\{
\begin{array}{ll}
\frac{\lambda^4 R^8  f^2}{96\pi^2} \left( \frac{\lambda }{\lambda_0} \right)^2|Z_L|^2 &: \lambda  R^2 |Z_L| \lesssim   \lambda_0 f \\
\frac{\lambda^4 R^4 f^4}{16\pi^2} {\rm ln} \frac{\lambda R^2  |Z_L|}{ \lambda_0 f } & : \lambda R^2  |Z_L| \gtrsim  \lambda_0 f,
\end{array}
\right. \nonumber \\
f(x) =& (4 + x^2)^{-2} \Bigl[
32+ 20 x^2 + 3 x^4   \nonumber \\
& + \left(  16 - 4 \sqrt{1 + 4 /x^2} + 8x^2 + x^4 - 6 x \sqrt{4 + x^2} \right. \nonumber \\
 &\left. - x^3 \sqrt{4 + x^2} \right) {\rm ln} \left( 1 + \frac{x^2}{2} - x \sqrt{1 + x^2/4} \right) \nonumber \\
&
+ \left(  16 + 4 \sqrt{1 + 4 /x^2} + 8x^2 + x^4 + 6 x \sqrt{4 + x^2} \right. \nonumber \\
&  \left.+ x^3 \sqrt{4 + x^2} \right) {\rm ln} \left( 1 + \frac{x^2}{2} + x \sqrt{1 + x^2/4}\right) 
\Bigr]. \nonumber
\end{align}
In this model, $m_{Z}^2$ is given by
\begin{align}
\label{eq:mZ_IYIT}
m_Z^2 = \frac{\lambda^4 R^8  f^2}{96\pi^2} \left( \frac{\lambda }{\lambda_0} \right)^2 + \frac{1}{2}\frac{\partial^2 \Delta V_\pm}{\partial |\vev{Z_L}|^2}.
\end{align}

\subsection{Parameter window of the IYIT model}
Let us now discuss constraints on the parameter space.
The constraint from the stability of the vacuum, $\lambda_0' v^2 > \lambda f^2$, is
\begin{align}
\label{eq:stability}
\lambda R < \lambda_0.
\end{align}
Constants $\lambda'r$, $\lambda'/r$ and $\lambda_0'$ are dimensionless coupling constants in the IYIT model, and are at the most $O(1)$.
This gives upper bounds on $\lambda_0$ and $R$,
\begin{align}
\label{eq:coupling1}
\lambda R^3 <1,\\
\label{eq:coupling2}
\lambda_0 R <1.
\end{align}
Finally, the potential of $Z_L$ becomes logarithmic for $\lambda R^2 Z_L > \lambda_0 f$, and cannot stabilize the sgoldstino against the tadpole term, so
\begin{align}
\label{eq:field value}
\lambda R < \sqrt{ \frac{4\pi}{\sqrt{2}\alpha_3} \frac{m_{\tilde{g}}}{f} } \lambda_0^{1/2}.
\end{align}
By combining the bounds in Eqs.~(\ref{eq:stability}-\ref{eq:field value}), we obtain upper bounds on $R^4 / \lambda_0$,
\begin{align}
\label{eq:bound_mZ}
\frac{R^4}{\lambda_0} < 
 \left\{
\begin{array}{ll}
\lambda^{-2} & {\rm Eqs.}~(\ref{eq:stability}), (\ref{eq:coupling1}) \\
\lambda^{-8/3} h^2 & {\rm Eqs.}~(\ref{eq:coupling1}), (\ref{eq:field value}) \\
\lambda^{-10/3} h^{10/3} & {\rm Eqs.}~(\ref{eq:coupling2}), (\ref{eq:field value}) 
\end{array}
\right.,~~
h \equiv \sqrt{ \frac{4\pi}{\sqrt{2} \alpha_3} \frac{m_{\tilde g}}{f} }
\end{align}
These give upper bounds on $m_Z^2$.

\begin{figure}[tb]
 \begin{center}
  \includegraphics[width=0.8\linewidth]{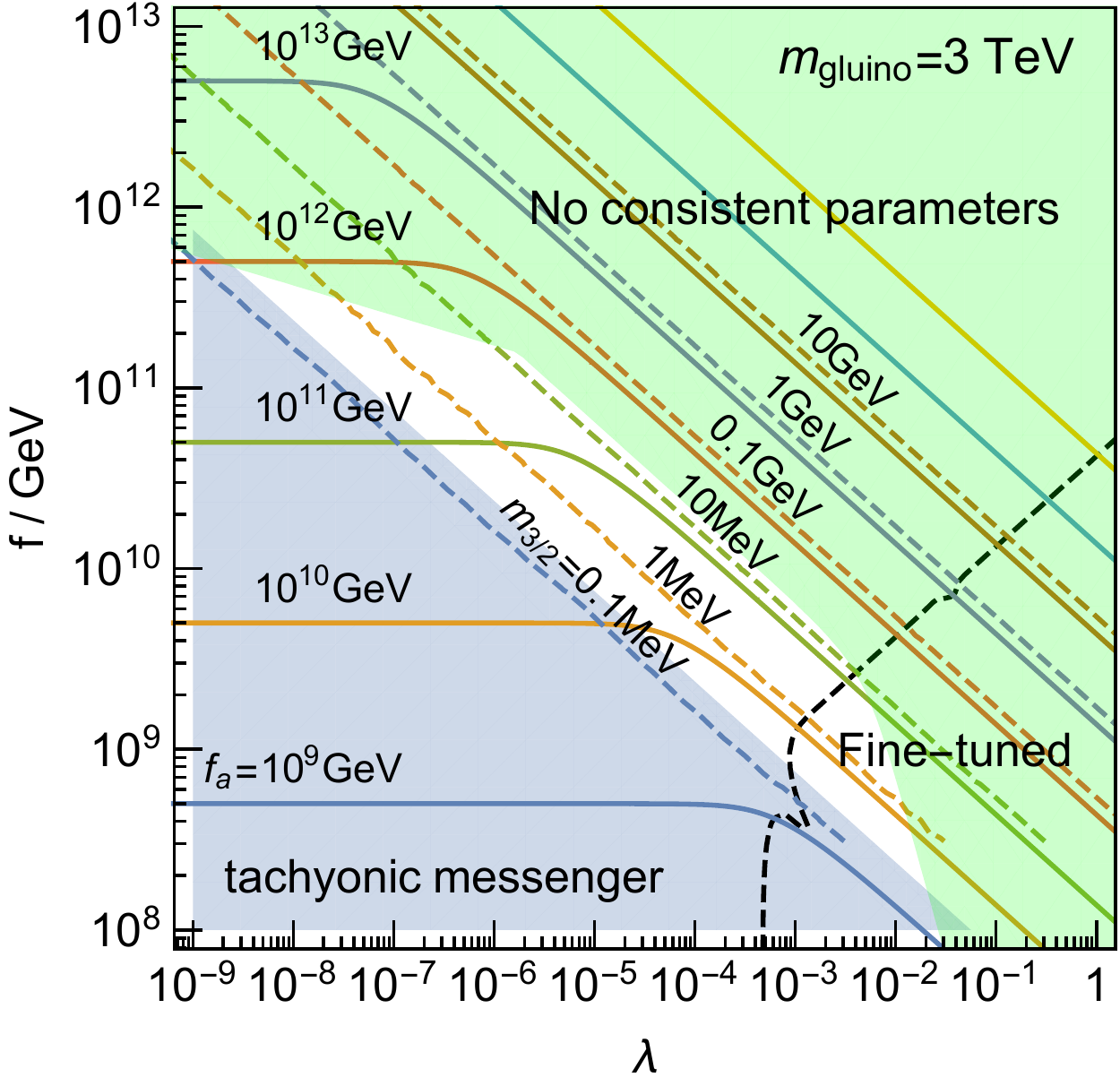}
 \end{center}
\caption{\small\sl 
The bounds on $(\lambda,f)$ for the IYIT model and the contours of the axion decay constant $f_a$ and the gravitino mass $m_{3/2}$.
The blue shaded region is excluded as the messenger field is tachyonic.
The region below the black dashed line requires fine-tuning.
There is no consistent parameter $(\lambda_0,R)$ to yield the green shaded region.
}
\label{fig:constraint_IYIT}
\end{figure}

In Fig.~\ref{fig:constraint_IYIT}, we show the constraints on $(\lambda,f)$.
The meaning of the blue shaded region and the black dashed line are the same as in Fig.~\ref{fig:constraint_ind}.
In the green shaded region, the bound on $m_Z^2$ from Eqs.~(\ref{eq:mZ_IYIT}) and (\ref{eq:bound_mZ}) is inconsistent with the required value of $m_Z^2$ shown in Eq.~(\ref{eq:mZ_required}).
We obtain the bounds on the axion decay constant $f_a$ and the gravitino mass $m_{3/2}$ 
\begin{align}
10^9~{\rm GeV} \lesssim f_a \lesssim 10^{12}~{\rm GeV},\\
0.1~{\rm MeV} \lesssim m_{3/2} \lesssim 10~{\rm MeV},
\end{align}
 for a gluino mass $\mathcal{O}$(TeV). It is interesting that the allowed range of $f_a$ is consistent with the axion dark matter scenario~\cite{Preskill:1982cy,Abbott:1982af,Dine:1982ah,Davis:1986xc}.

\section{Summary}
In this letter, we have presented a model that tackles several outstanding issues in the Standard Model and its supersymmetric extension.

We have examined a minimal hidden sector that consists of a superpotential with a $U(1)$ symmetry, which we identify with the PQ symmetry, and messenger quarks that carry $SU(3)_c$ charges.  
 Supersymmetry and this PQ symmetry are spontaneously broken while lowest order supergravity effects create the messenger scale.  Quantum effects generate a potential for the sgoldstino and force constraints on model parameters to ensure the stability of the SUSY-breaking vacuum. These constraints proved to be too stringent and required that we supplement the minimal model with extra matter fields. We have shown that classes of models that share features with ours, such as a quantum mechanically induced sgoldstino mass and a minimal messenger sector, automatically obtain lower bounds on the axion decay constant and gravitino mass. This fact encouraged us to supplement our minimal model in the hopes that such attractive features could be preserved and expanded upon in a stable extended model.

An IYIT model with $SU(2)$ gauge dynamics is a natural candidate for such an extended model since the minimal model is easily embedded in the $U(1)$ charged subsector of this larger model. Combining the inequalities from vacuum stability and IYIT coupling constants, upper bounds for the sgoldstino mass were derived. The resulting window in parameter space was found to restrict the gravitino mass to lie between 0.1 MeV $\lesssim m_{3/2}\lesssim$ 10 MeV  and the axion decay constant to $10^9$ GeV $\lesssim f_a \lesssim 10^{12}$~GeV, which is the suitable range for invisible axion dark matter.
 
\section*{Acknowledgement}
We thank Yasunori Nomura for collaboration in the early state of this work.  This work was supported in part by the Director, Office of Science, Office of High Energy and Nuclear Physics, of the U.S.\ Department of Energy under Contract DE-AC02-05CH11231, by the National Science Foundation under grants PHY-1316783 and PHY-1521446.

  \end{document}